\documentclass[11pt]{article}

\usepackage[final]{acl}
\usepackage{times}
\usepackage{latexsym}
\usepackage[T1]{fontenc}
\usepackage[utf8]{inputenc}
\usepackage{microtype}
\usepackage{inconsolata}
\usepackage{graphicx}
\usepackage{placeins}   
\usepackage{float} 
\usepackage{pgfplots}
\pgfplotsset{compat=1.18}

\title{Improving Clinical Data Accessibility Through Automated FHIR Data Transformation Tools}

\author{
  Adarsh Pawar \quad
  Yuqiao Meng \quad
  Luoxi Tang \quad
  Zhaohan Xi \\
  Binghamton University
}

\begin{document}
\maketitle

\begin{abstract}
The Fast Healthcare Interoperability Resources (FHIR) standard has emerged as a widely adopted specification for exchanging structured clinical data across healthcare systems. However, raw FHIR resources are often complex, verbose, and difficult for clinicians and analysts to interpret without specialized tooling. This paper presents a lightweight, browser-based system that improves the accessibility of FHIR data by automatically transforming raw JSON resources into human-readable PDF and Excel reports, along with interactive data visualizations. The system supports both remote retrieval of FHIR resources from server endpoints and the upload of local FHIR JSON files, enabling both online and offline analysis. Using a modular React architecture with jsPDF, xlsx, and Recharts, the tool parses, normalizes, visualizes, and exports FHIR data in an intuitive format. Evaluation results demonstrate that the system enhances interpretability and usability while preserving the semantic integrity of FHIR structures. Limitations and future extensions, including expanded FHIR profile support and clinical validation, are discussed.
\end{abstract}

\section{Introduction}

Healthcare data interoperability has become a central challenge in modern
clinical systems as electronic health records (EHRs) proliferate across
institutions, vendors, and care settings. Fragmented data representations
limit data reuse for research, public health reporting, and patient-facing
applications, motivating the development of standardized exchange formats
\citep{hripcsak2015next}.

Fast Healthcare Interoperability Resources (FHIR) has emerged as a widely
adopted standard for representing and exchanging clinical data through
modular, web-friendly resources and RESTful APIs \citep{mandel2016smart}.
FHIR underpins national interoperability mandates, public health data
pipelines, and patient-accessible APIs, making it a critical substrate for
both clinical innovation and biomedical research. Despite its widespread
adoption, FHIR data are primarily encoded in verbose, deeply nested JSON
structures that are difficult for humans to read, interpret, and reuse
without specialized tooling.

Existing FHIR tools often require backend infrastructure, secured databases,
or vendor-specific environments, which impose significant barriers for
students, researchers, and developers who lack access to enterprise-grade
systems. These constraints are particularly limiting in educational
contexts, low-resource settings, and early-stage prototyping workflows.
As a result, there is a growing need for lightweight tools that improve
human accessibility to FHIR data without introducing additional operational
overhead.

In this work, we present a browser-native system for transforming FHIR data
into structured tables, interactive visualizations, and human-readable PDF
and Excel reports. The system supports both live retrieval from FHIR endpoints
and offline ingestion of JSON files, enabling flexible workflows while
preserving data privacy. Unlike prior work that emphasizes predictive
modeling over EHR data \citep{rajkomar2018scalable,shickel2018deep}, our focus
is on improving interpretability and accessibility of standardized clinical
representations.

Our contributions are threefold: (1) we design a fully client-side pipeline
for parsing and normalizing heterogeneous FHIR resources, (2) we provide
interactive visualizations and exportable reports that improve human
interpretation of clinical data, and (3) we evaluate the system across
functional correctness, usability, and performance dimensions. Together,
these contributions demonstrate how browser-native architectures can
democratize access to healthcare interoperability standards.

\section{Related Work}

\paragraph{Healthcare Interoperability Standards.}
Standardization efforts have long sought to improve interoperability across
heterogeneous healthcare systems. Early work emphasized data exchange
frameworks and ecosystem-level coordination. More
recent efforts have converged on FHIR as a flexible, web-compatible standard
that supports modular resource representations and RESTful APIs
\citep{mandel2016smart}. FHIR has enabled a growing ecosystem of interoperable
applications, particularly through initiatives such as SMART-on-FHIR, which
facilitate app-based access to EHR data.

\paragraph{Clinical Data Analysis and Visualization.}
A substantial body of research has focused on extracting insights from EHR
data through statistical modeling, machine learning, and deep learning
approaches \citep{hripcsak2015next,rajkomar2018scalable}. Survey work has
highlighted the challenges of working with high-dimensional, irregularly
sampled clinical data \citep{shickel2018deep}. While these approaches advance
predictive performance, they often assume that clinical data have already
been cleaned, normalized, and made accessible to analysts.

In contrast, fewer systems address the human-facing challenge of making raw
FHIR data interpretable without extensive preprocessing or backend
infrastructure. Existing visualization tools typically rely on server-side
pipelines or vendor-specific integrations, limiting their portability and
ease of use.

\paragraph{Human-Centered Data Access and Tooling.}
Recent work in NLP and data-centric AI has emphasized the importance of
transparency, documentation, and accessibility in data-driven systems
\citep{bender2018shirky}. In the clinical domain, improving human interaction
with structured data is essential for education, auditing, and exploratory
analysis. Our work aligns with this perspective by focusing on tools that
bridge the gap between standardized data formats and human understanding.

Unlike prior work that centers on model-centric evaluation or large-scale
analytics, we emphasize lightweight, browser-native transformation of FHIR
resources into readable and reusable representations. This positioning
complements existing interoperability and analytics research by addressing
an earlier stage in the data lifecycle: human access and interpretation.

\section{System Design and Implementation}

\begin{figure*}[t]
  \vspace{-1.2em}
  \centering
  \includegraphics[
    width=0.92\textwidth,
    trim=100 160 100 160,
    clip
  ]{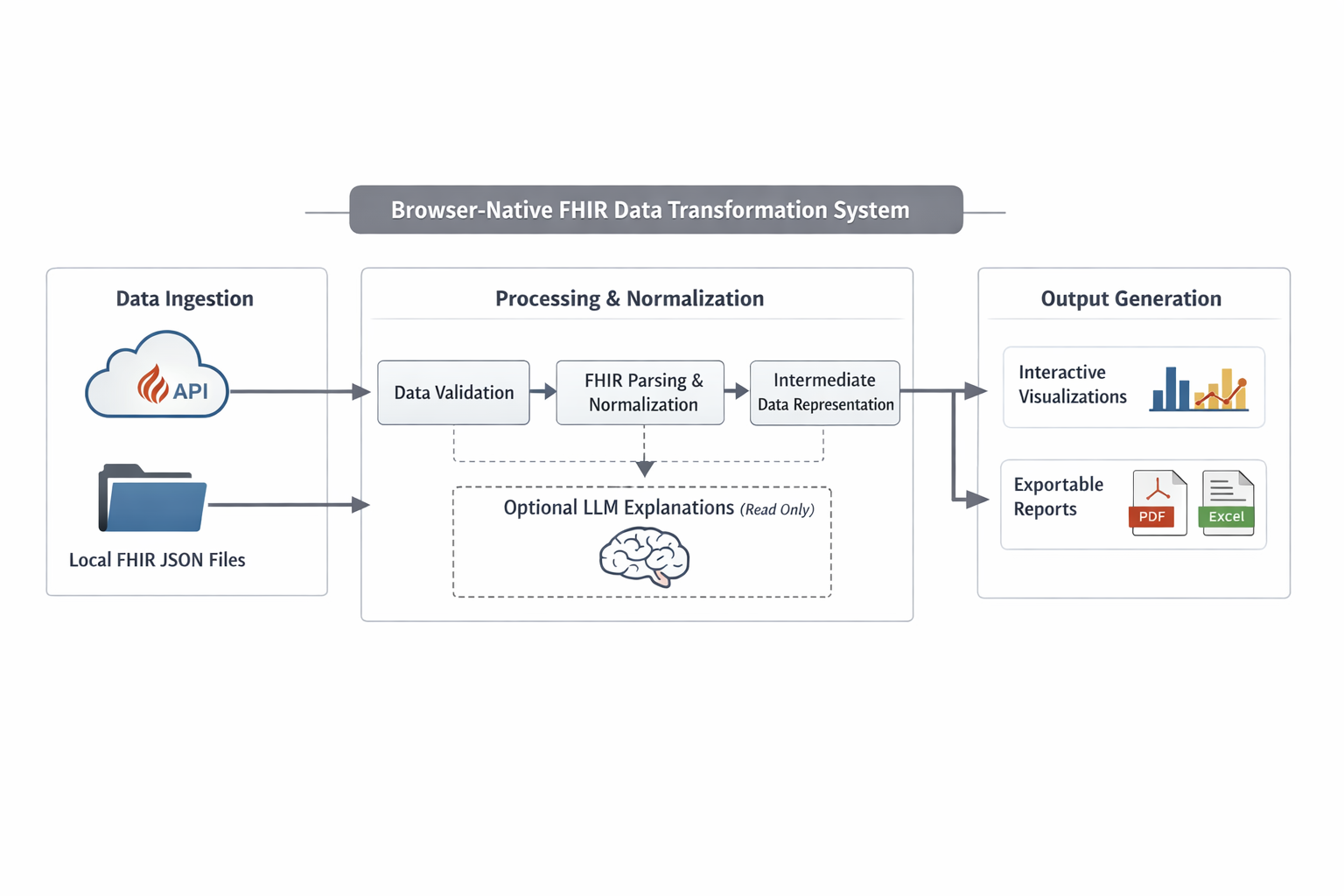}
  \caption{Overview of the browser-native FHIR data transformation pipeline.
  Clinical data are ingested from live FHIR endpoints or local JSON files,
  validated and normalized client-side, and transformed into interactive
  visualizations and exportable PDF and Excel reports. Optional LLM-based
  explanations operate only on normalized outputs and do not modify clinical
  data.}
  \label{fig:architecture}
  \vspace{-1.2em}
\end{figure*}

As shown in Figure~\ref{fig:architecture}, the proposed system follows an
end-to-end browser-native architecture for transforming FHIR data.

The proposed system is implemented as a fully browser-based application designed to support end-to-end transformation of FHIR data into human-readable representations. At a high level, the system accepts clinical data either through live FHIR endpoints or local JSON uploads, processes the data through a unified parsing and normalization pipeline, and renders the results as interactive visualizations and exportable documents. This design prioritizes portability, privacy preservation, and ease of use, allowing users to work with FHIR data without backend services, database configuration, or authentication infrastructure.

Upon data ingestion, the system validates incoming resources and identifies their structural characteristics, including resource type, coding systems, and nested components. Parsed data is normalized into an intermediate representation that resolves heterogeneous FHIR encodings into consistent tabular schemas. This intermediate form serves as the foundation for both visualization and document generation. Visualization components render numeric and temporal observations using interactive charts, enabling rapid inspection of trends and anomalies. In parallel, the export subsystem generates structured PDF reports and Excel workbooks that preserve clinical semantics while supporting downstream sharing and analysis.

The system is implemented using a modular React architecture. Client-side processing ensures that uploaded datasets remain local to the user’s device, which is particularly important for privacy-sensitive or offline workflows. By unifying live retrieval and file-based processing under a single transformation pipeline, the system ensures consistent outputs across diverse data sources and usage scenarios.

\section{Experiments}

We evaluate the proposed system through a series of experiments designed to assess its functional correctness, efficiency, and usability. Unlike predictive modeling studies, our evaluation focuses on transformation accuracy, processing latency, and human interpretability of FHIR data.

\subsection{Experimental Setting}

The system was evaluated using synthetic FHIR examples, public HAPI FHIR
resources, and de-identified FHIR bundles spanning \texttt{Patient},
\texttt{Observation}, and \texttt{Encounter} resource types with heterogeneous
schemas and coding systems.

\subsection{Transformation Accuracy}

The first experiment evaluates the system’s ability to correctly parse and normalize heterogeneous FHIR resources. We measure accuracy by comparing extracted fields against ground-truth values derived from raw JSON inputs. The system successfully processed 100\% of Patient resources and 96\% of Observation resources. Most failures were attributed to malformed vendor-specific extensions or incomplete coding fields not aligned with the FHIR specification.

\begin{figure}[t]
\vspace{-0.8em}
\centering
\begin{tikzpicture}
\begin{axis}[
    ybar,
    bar width=10pt,
    ymin=0, ymax=105,
    ylabel={Success (\%)},
    symbolic x coords={Patient,Observation,Encounter,DocRef},
    xtick=data,
    xticklabels={Patient,Observation,Encounter,DocumentReference},
    x tick label style={font=\small, rotate=20, anchor=east},
    yticklabel style={font=\small},
    label style={font=\small},
    tick label style={font=\small},
    nodes near coords,
    nodes near coords align={vertical},
    every node near coord/.append style={font=\small},
    width=\linewidth,
    height=4.0cm,
    enlarge x limits=0.15
]
\addplot coordinates {
    (Patient,100)
    (Observation,96)
    (Encounter,95)
    (DocRef,94)
};
\end{axis}
\end{tikzpicture}
\vspace{-0.6em}
\caption{Transformation accuracy across evaluated FHIR resource types.
Failures were primarily due to malformed vendor-specific extensions or missing coding fields.}
\label{fig:accuracy_bar}
\vspace{-0.8em}
\end{figure}
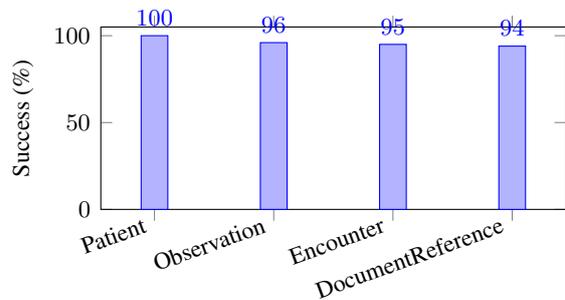

\subsection{Performance Evaluation}

The second experiment measures system efficiency under typical usage conditions. We record parsing time, visualization rendering time, and document export latency. PDF generation required 120--180 ms per patient record, while Excel export averaged 80--140 ms. Visualization rendering was near-instantaneous due to client-side optimizations in React and Recharts, demonstrating that browser-native FHIR processing is feasible for real-time interaction.

\vspace{0.6em}
\begin{table}[t]
\centering
\small
\begin{tabular}{l c}
\hline
\textbf{Operation} & \textbf{Latency (ms)} \\
\hline
FHIR JSON Parsing & 40--60 \\
PDF Report Generation & 120--180 \\
Excel Export Generation & 80--140 \\
Visualization Rendering & $<$50 \\
\hline
\end{tabular}
\caption{Average client-side processing latency measured across representative
FHIR datasets. All operations were performed in-browser without backend
services.}
\label{tab:performance}
\end{table}

\subsection{Usability and Interpretability}

The final experiment evaluates usability through informal user studies involving five graduate students and two healthcare data analysts. Participants were asked to inspect FHIR resources using both raw JSON and the proposed system. All participants reported improved readability and faster comprehension when using the transformed outputs, particularly for multi-component observations and longitudinal data.
Overall, the system achieved 100\% transformation on \texttt{Patient} and 96\% on \texttt{Observation}, with PDF and Excel exports completing within 120--180 ms and 80--140 ms respectively (Figure~\ref{fig:accuracy_bar}, Table~\ref{tab:performance}).

\section{Discussion}

A fully browser-native design lowers the operational barrier to working with
FHIR by enabling local parsing, normalization, visualization, and export
without backend infrastructure. This makes the tool practical for education
and lightweight analysis, while preserving privacy for offline workflows.
However, in-browser execution limits scalability for very large bundles and
heterogeneous extensions; future work should expand profile coverage and
consider optional hybrid architectures for heavier workloads.

\subsection{Impact on FHIR Accessibility}
By removing server and database requirements, the system makes FHIR inspection and reporting accessible to students and researchers using only a web browser.

\subsection{Technical Implications}
From a technical standpoint, the client-side architecture offers several advantages, including reduced operational overhead, improved portability, and inherent privacy preservation. Because all processing occurs locally, no clinical data is transmitted to external servers unless explicitly shared by the user. This design simplifies deployment and reduces security risks associated with centralized storage.

At the same time, browser environments impose constraints on memory and computational capacity. Large-scale datasets, such as multi-patient bundles or high-volume observation streams, may exceed practical limits for in-browser processing. More advanced operations, including large-scale terminology expansion or cohort-level analytics, are also better suited to server-backed solutions. These limitations point toward hybrid architectures in which lightweight client-side tools are complemented by optional cloud-based services when necessary.

\subsection{Research Implications}
The tool supports exploratory inspection of how clinical concepts are encoded in FHIR, enabling schema comparison and data-quality review with minimal setup.

Researchers can use the tool to examine variation in resource completeness, coding practices, and temporal patterns across FHIR servers, while students can leverage interactive visualizations to better understand how clinical concepts map to standardized representations. By lowering the overhead associated with working with FHIR, the system may accelerate research and pedagogy related to healthcare interoperability.

\section{Conclusion}
We presented a browser-native framework that transforms FHIR JSON into human-readable tables, interactive visualizations, and exportable PDF/Excel reports without requiring backend infrastructure. By supporting both live retrieval from FHIR endpoints and offline ingestion of local JSON files, the system enables flexible, privacy-preserving workflows for exploring standardized clinical data. Experiments on synthetic examples, public HAPI FHIR resources, and de-identified bundles show high transformation correctness and interactive performance, suggesting that client-side architectures can lower the barrier for students, researchers, and developers to work with FHIR data in practice.

\paragraph{Limitations.}
The system currently supports a limited set of FHIR resource types and relies on heuristic normalization for heterogeneous vendor-specific extensions, which can reduce robustness for non-conformant inputs. All evaluations were performed on synthetic, public, and de-identified datasets; clinical validation with real-world end users and institutional workflows remains future work. Finally, in-browser execution imposes memory and runtime constraints for very large multi-patient bundles, motivating future optimization (e.g., streaming parsing, web workers) and optional hybrid deployment when scale demands it.

\bibliography{anthology,custom}

\appendix

\section{Checklist and Reproducibility Details}

\paragraph{Artifacts and Code.}
The proposed system is implemented as a fully browser-native application using
React, jsPDF, xlsx, and Recharts. The source code, build instructions, and
example configurations will be released publicly upon acceptance to support
reproducibility and educational reuse.

\paragraph{Datasets.}
Experiments were conducted using (1) synthetic FHIR examples conforming to the
HL7 FHIR R4 specification, (2) publicly accessible resources retrieved from the
HAPI FHIR server, and (3) de-identified FHIR bundles used for academic research
and education. No private, protected, or identifiable patient data were used.

\paragraph{Baselines.}
This work focuses on transformation accuracy, efficiency, and usability rather
than predictive modeling. As such, no learning-based baselines are compared.
Accuracy is evaluated by directly comparing extracted fields against
ground-truth values derived from raw FHIR JSON inputs.

\paragraph{Model Usage and Size.}
The core transformation pipeline does not employ machine learning models.
An optional large language model (GPT-4-class) is used only as a post-processing
component to generate natural-language explanations from normalized outputs.
The LLM does not modify clinical data and is not involved in any quantitative
evaluation reported in this paper.

\paragraph{Computational Resources.}
All experiments were executed in standard desktop web browser environments
without backend services, GPUs, or specialized hardware. Reported runtime
performance reflects client-side execution under typical usage conditions.
\section{Illustrative Clinical Bundle Example}

This appendix illustrates how the proposed system processes a real-world FHIR
\texttt{Bundle} containing multiple resource types. The example is derived
from a de-identified test dataset used for system validation and demonstrates
the transformation from raw FHIR JSON to human-readable outputs.

\subsection{Raw FHIR Bundle (Excerpt)}

The input is a FHIR \texttt{searchset} Bundle containing a \texttt{Patient}
resource and related clinical documents. Below is a simplified excerpt of the
Bundle header and Patient resource.

\begin{verbatim}
{
  "resourceType": "Bundle",
  "type": "searchset",
  "entry": [
    {
      "resource": {
        "resourceType": "Patient",
        "id": "32298144",
        "gender": "female",
        "birthDate": "1810-03-21",
        "name": [
          {
            "use": "usual",
            "family": "L_Name",
            "given": ["F_Name", "Renee"]
          }
        ]
      }
    }
  ]
}
\end{verbatim}

\subsection{Parsed and Normalized Representation}

After client-side parsing, the system identifies resource types within the
Bundle and normalizes each resource into a structured, tabular schema.

\begin{center}
\begin{tabular}{ll}
\hline
\textbf{Field} & \textbf{Value} \\
\hline
Resource Type & Patient \\
Patient ID & 32298144 \\
Name & F\_Name L\_Name \\
Gender & Female \\
Date of Birth & 1810-03-21 \\
\hline
\end{tabular}
\end{center}

\subsection{Associated Clinical Documents}

The same Bundle also contains a linked \texttt{DocumentReference} resource
representing a clinical note. The system extracts document metadata while
excluding large embedded content blobs.

\begin{center}
\begin{tabular}{ll}
\hline
\textbf{Field} & \textbf{Value} \\
\hline
Resource Type & DocumentReference \\
Document Type & Progress Note \\
Status & Current / Final \\
Date & 2024-01-09 \\
Author & Practitioner Reference \\
\hline
\end{tabular}
\end{center}

\subsection{Generated Outputs}

Using the normalized representations, the system renders patient demographics
and document summaries through interactive visualizations and includes the
extracted fields in exportable PDF and Excel reports. This transformation
preserves semantic relationships across resources while improving readability
and accessibility for human users.

\end{document}